\documentclass[aps,prl,twocolumn,pacs]{revtex4-1}
\usepackage{amsmath,bm,epsfig}
\usepackage{epsfig}
\usepackage{bm}
\newcommand{\B}[1]{{\bm{#1}}}
\newcommand{\C}[1]{{\mathcal{#1}}}
\newcommand{\pa}{\partial}
\usepackage[latin1]{inputenc}

\begin{document}
\title{Autonomy and Singularity in Dynamic Fracture}
\author{Eran Bouchbinder}
\affiliation{Department of Chemical Physics, Weizmann Institute of Science, Rehovot 76100, Israel}
%\date{\today}
\begin{abstract}
The recently developed weakly nonlinear theory of dynamic fracture predicts $1/r$ corrections to the standard asymptotic linear elastic $1/\sqrt{r}$ displacement-gradients, where $r$ is measured from the tip of a tensile crack. We show that the $1/r$ singularity does not automatically conform with the notion of autonomy (autonomy means that any crack tip nonlinear solution is uniquely determined by the surrounding linear elastic $1/\sqrt{r}$ fields) and that it does not automatically satisfy the resultant Newton's equation in the crack parallel direction. We show that these two properties are interrelated and that by requiring that the resultant Newton's equation is satisfied, autonomy of the $1/r$ singular solution is retained. We further show that the resultant linear momentum carried by the $1/r$ singular fields vanishes identically. Our results, which reveal the physical and mathematical nature of the new solution, are in favorable agreement with recent near tip measurements.
\end{abstract}
\maketitle

A weakly nonlinear theory of dynamic fracture, which extends the standard theory of fracture \cite{98Fre, 99Bro}, was recently developed \cite{08BLF, 09BLF}. This theory was shown to be in excellent agreement with groundbreaking experimental measurements of the deformation near the tip of rapid cracks \cite{08LBF, 08BLF, 09BLF, 10LBSF}. Furthermore, it may be relevant to understanding currently poorly understood crack tip instabilities \cite{09Bouchbinder}. In this Rapid Communication we derive a series of theoretical results that further elucidate the physical and mathematical nature of the theory.

The standard approach to dynamic fracture - linear elastic fracture mechanics (LEFM) \cite{98Fre, 99Bro} - assumes that linear elasticity dominates the deformation fields outside a small region near the tip of a crack. Its major prediction is that crack tips concentrate large deformation-gradients and stresses which are characterized by a universal $1/\sqrt{r}$ singular behavior. Many results in this theoretical framework are derived from the latter property \cite{98Fre, 99Bro}.

The basic physical idea underlying the weakly nonlinear theory of dynamic fracture is that linear elasticity breaks down when elastic nonlinearities intervene near the tip of a crack. This is a physically intuitive idea since atomes/molecules are expected to sample reversible (i.e. elastic) anharmonic parts of the interaction potential before their separation is large enough to induce irreversible deformation (e.g. damage, plasticity and eventually fracture).

To mathematically formulate this idea, we consider the following expansion of the displacement field $\B u$ \cite{08BLF, 09BLF}
\begin{eqnarray}
\label{expansion}
\B u(r,\theta;v) &\simeq& \epsilon  \tilde{\B u}^{(1)}(r,\theta;v)+\epsilon^2 \tilde{\B u}^{(2)}(r,\theta;v)+ \C O(\epsilon^3)\nonumber\\
 &\equiv& \B u^{(1)}(r,\theta;v) + \B u^{(2)}(r,\theta;v)+ \C O(\epsilon^3)\ ,
\end{eqnarray}
where $\epsilon$ quantifies the magnitude of the displacement-gradients \cite{08BLF}, $(r,\theta)$ is a polar coordinates system located at a crack's tip and moving with it at a speed $v$ in the $\theta\!=\!0$ direction. The first order term in $\epsilon$ corresponds to linear elasticity, which is actually only a first term in a more general expansion, while the second order term corresponds to the leading nonlinearity that intervenes when the deformation is large enough. As higher order nonlinearities are neglected in Eq. (\ref{expansion}), the theory based on it is termed ``weakly nonlinear theory of dynamic fracture'' \cite{08BLF, 09BLF, 08BL}.

The expansion in Eq. (\ref{expansion}) can be substituted in a general elastic strain energy functional $U(\B F)$, where $F_{ij}\!=\!\delta_{ij}+\pa_j u_i$, from which the first Piola-Kirchhoff stress tensor $\B s$ can be derived as
\begin{equation}
\label{1st_PK}
\B s = \frac{\pa U}{\pa \B F} \ .
\end{equation}
$\B s$ quantifies forces in the deformed configuration per unit areas in the reference configuration \cite{Holzapfel}.
Then, the momentum balance equations and the crack faces traction-free boundary conditions are obtained order by order in $\epsilon$ \cite{08BLF, 09BLF, 08BL}. $\B u^{(1)}(r,\theta;v)$ in Eq. (\ref{expansion}) satisfies the first order problem, which is a standard LEFM one \cite{98Fre, 99Bro}. The near crack tip (asymptotic) fields for steady state propagation under Mode I (opening) symmetry are given by \cite{98Fre, 99Bro}
\begin{eqnarray}
u_x^{(1)}(r, \theta;v)&=&\frac{K_I \sqrt{r}}{4\mu\sqrt{2\pi}}\Omega_x(\theta;v),\nonumber\\
\label{firstO}
u_y^{(1)}(r,\theta;v)&=&\frac{K_I\sqrt{r}}{4\mu\sqrt{2\pi}}\Omega_y(\theta;v).
\end{eqnarray}
Here $K_I$ is the Mode I ``stress intensity factor'' which
cannot be determined by the asymptotic analysis, but rather by
the {\em global} crack problem. $\B \Omega(\theta;v)$ is a known
universal function \cite{98Fre, 99Bro, 08BLF} and $\mu$ is the shear modulus. $x$ corresponds to the propagation direction, $\theta\!=\!0$, and $y$ corresponds to the direction in which the tensile loadings are applied, $\theta \!=\! \pm \pi/2$. The displacement fields in Eq. (\ref{firstO}) give rise to the famous $1/\sqrt{r}$ displacement-gradients and stress singularity \cite{98Fre, 99Bro}.

$\B u^{(2)}\!(r,\theta;v)$ in Eq. (\ref{expansion}) satisfies the second order problem, which was explicitly derived in \cite{08BLF, 09BLF}. It has the following form
\begin{eqnarray}
\label{solution}
u_x^{(2)}(r,\theta;v)\!&=&\!\frac{K_I^2}{32\pi\mu^2}\Big[A\log{r}+\frac{A}{2}\log{\left(1-\frac{v^2\sin^2\theta}{c_d^2} \right)}\nonumber\\
+\,B\alpha_s\log{r}\!\!&+&\!\!\frac{B \alpha_s}{2}\log{\left(1-\frac{v^2\sin^2\theta}{c_s^2} \right)}+\Upsilon_x(\theta;v)\Big],\nonumber\\
u_y^{(2)}(r,\theta;v)\!&=&\!\frac{K_I^2}{32\pi\mu^2}\Big[-A\alpha_d\theta_d-B\theta_s+\Upsilon_y(\theta;v)\Big],
\nonumber\\
\tan{\theta_{d,s}}&=&\alpha_{d,s}\tan{\theta},\quad\alpha^2_{d,s}\equiv1-v^2/c_{d,s}^2 \ ,
\end{eqnarray}
where $c_{d,s}$ are the dilatational and shear wave speeds, respectively.
$\B \Upsilon(\theta;v)$ is given in the form
\begin{equation}
\Upsilon_x(\theta;v) \!=\! \sum_{n}\!\!c_n(v) \cos(n\theta),~~\Upsilon_y(\theta;v) \!=\! \sum_{n}\!\!d_n(v) \sin(n\theta),
\end{equation}
where the coefficients $\{c_n(v) ,d_n(v)\}$ can be easily obtained by solving a set of linear algebraic equations \cite{08BLF, 09BLF}.
The coefficients $A$ and $B$ are related by the traction-free boundary conditions on the crack faces \cite{08BLF,09BLF} through
\begin{equation}
A= \frac{2\mu B \alpha_s -(\lambda+2\mu)\pa_\theta \!\Upsilon_y(\pi;v)-\mu\,\kappa(v)}{\lambda - (\lambda+2\mu)\alpha_d^2} \ .
\label{A_B_bc}
\end{equation}
Here
\begin{eqnarray}
\label{kappa}
\kappa(v) = -\frac{16\alpha_d^2 v^4 \left(\lambda + \mu \right)}{\,\mu\, c_s^4 \left[4\alpha_s\alpha_d-(1+\alpha_s^2)^2 \right]^2}
\end{eqnarray}
and $\lambda$ is the second Lam\'e coefficient \cite{98Fre, 99Bro}. The displacement fields in Eq. (\ref{solution}) contain $\log{(r)}$ terms and give rise to $1/r$ singular displacement-gradients. Both of these features were directly verified by near tip measurements \cite{08LBF, 08BLF}.

 The solution in (\ref{solution})-(\ref{A_B_bc}) satisfies the second order momentum balance equations and the traction-free boundary conditions on the crack faces. Therefore, one may reach the conclusion that the second order solution contains a parameter $B$ that is not uniquely determined by the stress intensity factor $K_I$. If true, this result has profound theoretical implications as it suggests that the concept of the autonomy of the near crack tip nonlinear region \cite{98Fre, 99Bro} is not always valid. The basic idea behind this concept is that the mechanical state within the near tip nonlinear zone, which is surrounded by the LEFM fields of Eqs. (\ref{firstO}), is uniquely determined by the value of $K_I$ and is otherwise independent of the applied loadings and the geometric configuration in a given problem. This implies, for example, that systems with the same $K_I$, but with different applied loadings and geometric configurations, will be in the same mechanical state within the near tip nonlinear zone. Autonomy is a central concept in fracture mechanics \cite{98Fre, 99Bro}.

In \cite{08BLF} values of $B$ were directly extracted from the experimental data, without addressing the question of whether they can be in fact theoretically determined. In \cite{09BLF} it was shown that in the quasi-static limit, $v \!\to\! 0$, $B$ can be theoretically determined by $K_I$ and hence the autonomy of the near tip nonlinear region is retained. This may appear as a puzzling result, because Eq. (\ref{solution}), with Eq. (\ref{A_B_bc}), satisfies the asymptotic second order boundary-value problem for {\em all} $B$'s - what is then the missing physical ingredient that is not contained within the asymptotic boundary-value problem?

To answer this question, which was only partially addressed in \cite{09BLF}, consider the net force per unit sample thickness $\B f$ acting on a line of radius $r$ encircling a crack's tip
\begin{eqnarray}
\label{f}
f_i \equiv \int_{-\pi}^{\pi} s_{ij} n_j r d\theta \ ,
\end{eqnarray}
where $\B n$ is an outward unit normal on the circle. The  Mode I symmetry immediately implies $f_y\!=\!0$. Moreover, as no force is acting on the crack tip in the crack propagation direction (x), and focusing first on the quasi-static limit ($v \!\to\! 0$) in which material inertia does not play a role, we must have $f_x\!=\!0$.
In the framework of quasi-static LEFM, one can show that a $1/r$ contribution to $\B s$ results in $f_x\!\ne\!0$ \cite{74Rice}. Therefore, $f_x\!=\!0$ is satisfied within quasi-static LEFM if and only if the $1/r$ singularity is discarded altogether, i.e. its prefactor is set equal zero \cite{74Rice}. The $1/r$ singularity generates an unbalanced (spurious) force in the crack parallel direction (where no boundary conditions are imposed), even though it satisfies the asymptotic boundary-value problem. Hence, in the framework of LEFM this singularity is unphysical.

This conclusion is in sharp contrast to the corresponding situation in the quasi-static weakly nonlinear theory \cite{09BLF}. In this case, a $1/r$ singular contribution to $\B s$ also does not automatically lead to $f_x\!=\!0$, but the latter can be recovered without discarding the whole solution by properly choosing $B$ as a function of $K_I$ in Eqs. (\ref{solution}) and (\ref{A_B_bc}), retaining autonomy \cite{09BLF}. Therefore, the special property of the $1/r$ singularity discussed above is precisely the missing physical ingredient, which is not contained within the the asymptotic boundary-value problem, that ensures that autonomy is not violated. We thus see that in contrast to LEFM, the $1/r$ singularity in quasi-static weakly nonlinear fracture mechanics is a physically sound solution that does not violate any physical principle and conforms with the concept of autonomy.

The discussion above was restricted to the quasi-static limit. The generalization to the fully dynamic case, $v\!>\!0$, is somewhat more subtle because material inertia can play a role. To understand this, we write down the resultant Newton's equation for the material enclosed within a circle of radius $r$ around the tip. It is a balance between the force $\B f$ in Eq. (\ref{f}) and the time rate of change of linear momentum $\dot{\B p}$ (both per unit sample thickness)
\begin{eqnarray}
\label{f_dynamic}
\!\!\!\!\!\!f_i \equiv \int_{-\pi}^{\pi}\!\! s_{ij} n_j r d\theta = v^2 \rho \int_0^r \!\! r' dr' \int_{-\pi}^{\pi} \pa_{xx} u_i d\theta \equiv \dot{p}_i,
\end{eqnarray}
where the steady state relation $\pa_t\!=\!-v\pa_x$ was used and $r'$ is a dummy integration variable. We expect Eq. (\ref{f_dynamic}) to provide the necessary condition for determining $B$ in the dynamic case, though it is clear that $\B f\!\ne\!0$ is possible if $\dot{\B p} \ne 0$, without violating any physical law.

In order to demonstrate the latter possibility, consider the asymptotic first order solution given in Eq. (\ref{firstO}), which can be used to derive the standard first order (linear elastic) stress tensor $\B s^{(1)}$ \cite{98Fre}. Using $\B u^{(1)}$ and $\B s^{(1)}$ in Eq. (\ref{f_dynamic}), $\B f^{(1)}$ and $\dot{\B p}^{(1)}$ can be calculated. Recall that Mode I symmetry implies that $f_y\!=\!\dot p_y\!=\!0$, so all the discussion to follow focuses on the the crack parallel direction $x$.

%%%%%%% FIGURE 1 %%%%%%%%%%%%%%%%%%
\begin{figure}[here]
\centering
\epsfig{width=.4\textwidth,file=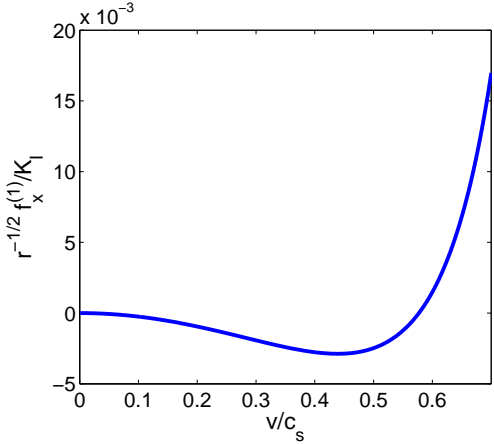}
\caption{(Color online) $r^{-1/2}f_x^{(1)}/K_I$ vs. $v/c_s$. An identical curve is obtained when $r^{-1/2}\dot{p}_x^{(1)}/K_I$ is plotted vs. $v/c_s$, which shows that $f_x^{(1)}=\dot p_x^{(1)}$.}\label{f_x}
\end{figure}
%%%%%%%%%%%%%%%%%%%%%%%%%%%%%%%%%%%

In Fig. \ref{f_x} we plot $r^{-1/2}f_x^{(1)}/K_I$ vs. $v/c_s$. The figure shows that $f_x^{(1)}\!\ne\! 0$ for $v\!>\!0$ (except for an isolated point). Moreover, a direct calculation shows that $f_x^{(1)}\!=\!\dot p_x^{(1)}$. Therefore, the standard LEFM $1/\sqrt{r}$ singularity automatically satisfies   Eq. (\ref{f_dynamic}), though $f_x^{(1)}\! \ne\! 0$. The situation is different when the $1/r$ singularity of the weakly nonlinear theory is considered.

Equation (\ref{f_dynamic}) was shown above to be satisfied automatically to first order in $\epsilon$, when the fields in Eq. (\ref{firstO}) are used. Consider now Eq. (\ref{f_dynamic}) to second order in $\epsilon$. The second order stress tensor $\B s^{(2)}$ has the following scaling property
\begin{equation}
s^{(2)} \sim \pa u^{(1)} \pa u^{(1)} + \pa u^{(2)} \sim r^{-1} \ ,
\end{equation}
where the tensorial notation was omitted for simplicity. This implies that $f_x^{(2)}$ is a constant independent of the radius $r$. On the other hand, we have
\begin{equation}
\pa_{xx}u^{(2)} \sim r^{-2} \ ,
\end{equation}
which implies that $\dot p_x^{(2)}$ depends on $r$ (in fact it diverge logarithmically). Therefore, the only way in which
Eq. (\ref{f_dynamic}) can be satisfied to second order is by having $\dot p_x^{(2)}\!=\!0$, which implies that
\begin{eqnarray}
\label{dynamic_condition}
f^{(2)}_x\equiv\int_{-\pi}^{\pi} s^{(2)}_{xj} n_j r d\theta = 0 \ .
\end{eqnarray}
Therefore, the $1/r$ singular fields carry no net linear momentum and $f_x^{(2)}\!=\!0$ is the condition that determines $B$, precisely as in the quasi-static limit.

The latter prediction, i.e. that $\dot p_x^{(2)}\!=\!0$, can be directly checked using the explicit second order solution in Eq. (\ref{solution}). In order to calculate $\dot p_x^{(2)}$, we evaluate the different contributions to $\pa_{xx} u_x^{(2)}$
\begin{widetext}
\begin{eqnarray}
&&\pa_{xx} \Upsilon_x(\theta;v) = \sum_{n} - n ~c_n(v) \sin\theta \frac{n \cos(n\theta) \sin\theta + 2\cos\theta \sin(n \theta)}{r^2} \ ,\nonumber\\
&&\pa_{xx} \left[A\log{r}+\frac{A}{2}\log{\left(1-\frac{v^2\sin^2\theta}{c_d^2} \right)}+B\alpha_s\log{r}+\frac{B \alpha_s}{2}\log{\left(1-\frac{v^2\sin^2\theta}{c_s^2} \right)}\right] =\nonumber\\
&& -\frac{2A c_d^2 \left[v^2+ \left(2c_d^2-v^2 \right)\cos(2\theta) \right]}{\left[2c_d^2 - v^2 + v^2 \cos(2\theta) \right]^2 r^2}
-\frac{2B c_s^2 \sqrt{1-v^2/c_s^2} \left[v^2+ \left(2c_s^2-v^2 \right)\cos(2\theta) \right]}{\left[2c_s^2 - v^2 + v^2\cos(2\theta) \right]^2 r^2} \ .
\end{eqnarray}
\end{widetext}
Hence, calculating analytically the angular integrals over the above expressions, which sum up to $\int_{-\pi}^{\pi} \pa_{xx} u_x^{(2)} d\theta$, we obtain
\begin{widetext}
\begin{eqnarray}
\!\!\!&&\int_{-\pi}^{\pi} \pa_{xx} \left[A\log{r}+\frac{A}{2}\log{\left(1-\frac{v^2\sin^2\theta}{c_d^2} \right)}+B\alpha_s\log{r}+\frac{B \alpha_s}{2}\log{\left(1-\frac{v^2\sin^2\theta}{c_s^2} \right)}+ \Upsilon_x(\theta;v) \right] d\theta=\\
\!\!\!&&\left[-\frac{A \sin(2\theta)}{\left[1+\alpha_d^2  + \frac{v^2}{c_d^2}\cos(2\theta) \right] r^2} -\frac{B  \alpha_s \sin(2\theta)}{\left[1 + \alpha_s^2+ \frac{v^2}{c_s^2}\cos(2\theta) \right] r^2}+ \sum_{n} n c_n(v) \frac{\sin[(n-2)\theta]-2\sin(n\theta)+\sin[(n+2)\theta]}{4r^2}\right]_{-\pi}^{\pi}\!\!\!=\!0 \nonumber \ .
\end{eqnarray}
\end{widetext}
which explicitly verifies that the $1/r$ singular fields carry no net linear momentum, $\dot p_x^{(2)}\!=\!0$.

%%%%%%% FIGURE 2 %%%%%%%%%%%%%%%%%%
\begin{figure}[here]
\centering
\epsfig{width=.4\textwidth,file=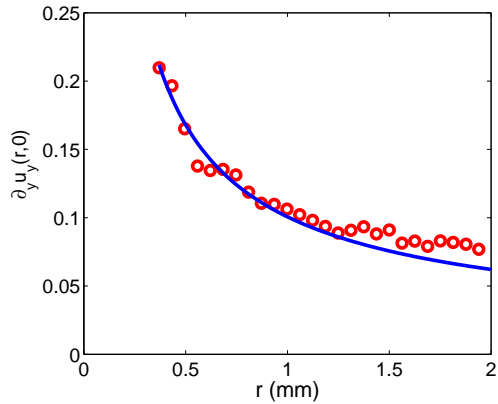}
\caption{(Color online) $\pa_y u_y(r,0)$ of the weakly nonlinear theory (solid line), cf. Eq. (\ref{expansion}), together with the experimental data (circles) of \cite{08LBF, 08BLF} for $v\!=\!0.53c_s$ and the material used in \cite{08LBF}. The theoretical curve is obtained without any fit, where only the stress intensity factor reported in Fig. 1(b) of \cite{08BLF} was used as an input, see text for more details.}\label{Strain_yy}
\end{figure}
%%%%%%%%%%%%%%%%%%%%%%%%%%%%%%%%%%%

We reiterate that surprisingly, even at finite crack velocities ($v\!>\!0$) inertia does not play a role in determining $B$ and in retaining autonomy, and the condition to be satisfied remains as in the quasi-static limit, i.e. Eq. (\ref{dynamic_condition}).
As was stressed several times above, Eq. (\ref{dynamic_condition}) is not satisfied for any $B$, but rather determines it. Therefore, in order to calculate $B$, we substitute $\B s^{(2)}$ (which is obtained from expanding Eq. (\ref{1st_PK}) in orders of $\epsilon$, cf. \cite{08BLF, 09BLF}) in the integrand of (\ref{dynamic_condition}) and look for the value of $B$ that makes the integral vanish. This cannot be done analytically, but is easily achieved numerically.

To test the theory, we compare its predictions to the direct near-tip deformation measurements of \cite{08LBF}. For that aim, we focus on $v\!=\!0.53c_s$ and use $\{c_n(0.53c_s) ,d_n(0.53c_s)\}$, $\lambda \!=\! 2\mu$, $\mu \!=\! 32.5$kPa as reported in \cite{08BLF} for the material used in \cite{08LBF}. Using these numbers in Eqs. (\ref{firstO})-(\ref{kappa}) to obtain $\B s^{(2)}$ and then calculating numerically the integral in Eq. (\ref{dynamic_condition}), we obtain $B \simeq 18.5$. This value cannot be directly compared to the value obtained by a fitting procedure in Fig. 1(b) of \cite{08BLF} because in the latter case the asymptotic LEFM fields of Eqs. (\ref{firstO}) had to be supplemented with a subleading term (corresponding to the ``T-stress'' \cite{08LBF, 08BLF}) and hence an additional parameter was involved. Instead, we focus on a smaller region near the tip, where the subleading term that we do not consider here is less significant, and use the stress intensity factor of Fig. 1(b) in \cite{08BLF}, $K_I \!=\! 1250$Pa$\sqrt{m}$. Recall that, as required by autonomy, the stress intensity factor is the only parameter that is needed as an input to the asymptotic near tip theory. In Fig. \ref{Strain_yy} we plot $\pa_y u_y(r,0)$ of the weakly nonlinear theory, cf. Eq. (\ref{expansion}), together with the experimental data of \cite{08LBF, 08BLF}. The agreement between the theory and the experimental data is remarkable, supporting the theoretical results derived above.

In summary, in this Rapid Communication we theoretically explored some properties of the $1/r$ singularity in the framework of the recently developed weakly nonlinear theory of dynamic fracture. It was shown that the theory is consistent with the notion of the autonomy of the nonlinear near-tip region for any crack tip velocity $v$, extending the quasi-static results of \cite{09BLF}. In addition, it was shown that no net linear momentum is carried by the $1/r$ singular fields. As only Mode I symmetry was considered here, a direction for future investigation is the development of a weakly nonlinear theory for more general fracture conditions.

\end{document}